# Unravelling the polarity of InN quantum dots using a modified approach of negative-spherical-aberration imaging†

Piu Rajak,[a] Mahabul Islam,[a,b] J. J. Jiménez,[c,d] J. M. Mánuel,[c,e] P. Aseev,‡[f] Ž. Gačević,[f] E. Calleja,[f] R. García,[c,d] Francisco M. Morales *[c,d] and Somnath Bhattacharyya

InN quantum dots (QDs) are considered to be promising nanostructures for different device applications. For any hexagonal AB-stacking semiconductor system, polarity is an important feature which affects the electronic properties. Therefore, the determination of this characteristic on any wurtzite (semi)polar III-N compound or alloy is essential for defining its applicability. In this paper, the polarity of InN QDs grown on silicon by indium droplet epitaxy plus nitridation and annealing was determined by a modified approach combining exit wave reconstruction with negative-spherical-aberration high-resolution lattice imaging using TEM. Comparing the micrographs of two QDs from the same TEM specimen with the simulated images of InN slab structures generated under the same conditions as of the experiments, it was confirmed that the QDs of the present study are N polar. Given that the settlement of material's polarity has always been a tedious, indirect and controversial issue, the major value of our proposal is to provide a straightforward procedure to determine the polar direction from atomic-resolution focal series images.

## 1. Introduction

InN is a technologically promising group III-V semiconductor compound due to its small energy bandgap (0.6–0.7 eV), high electron mobility,[1] small effective mass,[2] and high saturation velocity.[3] These properties make it a suitable candidate for optoelectronic devices, tunnel diodes, and high-frequency transistors.[4-8] Due to its surface selectivity towards certain solvents, InN becomes a potential candidate for chemical and biological sensors.[9-12] To further improve its applicability, InN has been used in the form of QDs in conjunction with other semiconductor alloy nanostructures.[13-16] An important aspect of any hexagonal AB-stacking semiconductor alloy system like InN is the polarity, which is related to both the non-centrosymmetry of the wurtzite unit cell and the partial ionicity of the atoms in the commonly observed dumbbells comprising one 'A' (cationic) and one 'B' (anionic) atom.[17] The 2D projected images of the oppositely charged atomic couples along the growth direction are commonly known as dumbbells. In a dumbbell, if 'A' is placed above 'B', it is called 'A' polar. Similarly, if 'B' is placed above 'A', it is called 'B' polar. In polar oriented materials, the polarity determines the defect concentration and doping nature, optical and optoelectronic properties, different chemical and physical behaviours, *etc*.[18] The changing polarity of a semiconductor material may result in higher electron mobility,[19,20] improved device performance[4-7,21] and enhanced detection sensitivity for hydrogen in dilute concentration.[22] Therefore, determining the polarity of InN QDs is essential to tune the material's properties for better device applications. To the best of our knowledge, until now there was no report available on the polarity determination of InN QDs.

The polarity is generally determined using different Transmission Electron Microscopy (TEM) techniques and X-ray photoelectron diffraction.[23] Among the TEM techniques, Convergent Beam Electron Diffraction (CBED)[24] and Scanning Transmission Electron microscopy (STEM) imaging[17,21,25,26] have been extensively used to determine the polarity of different hexagonal AB-stacking systems. In the case of CBED, simulated

[a]Department of Metallurgical and Materials Engineering, Indian Institute of Technology Madras, Chennai, 600036, India. E-mail: somnathb@iitm.ac.in
[b]Department of Physics, Indian Institute of Technology Madras, Chennai, 600036, India
[c]IMEYMAT: Institute of Research on Electron Microscopy and Materials, University of Cádiz, Spain. E-mail: fmiguel.morales@uca.es
[d]Department of Materials Science and Metallurgic Engineering, and Inorganic Chemistry, Faculty of Sciences, University of Cádiz, Puerto Real, 11510 Cádiz, Spain
[e]Department of Condensed Matter Physics, Faculty of Sciences, University of Cádiz, Puerto Real, 11510 Cádiz, Spain
[f]Instituto de Sistemas Optoelectrónicos y Microtecnología, Universidad Politécnica de Madrid, Ciudad Universitaria s/n, 28040 Madrid, Spain
†Electronic supplementary information (ESI) available. See DOI: 10.1039/c9nr04146j
‡Current address: QuTech and Kavli Institute of Nanoscience, Delft University of Technology, 2628 CJ Delft, The Netherlands

CBED patterns are matched with experimental CBED patterns with varying thicknesses. However, CBED is not suitable for quantum dots and nanostructures. STEM imaging combined with a probe corrector allows high-resolution Z-contrast (Z: atomic number) imaging, *i.e.* elements with a higher atomic number will have higher intensity in HAADF images. By comparing the experimental image intensity of the constituent elements with simulated images, the polarity can be determined. Exit wave reconstruction[27-30] and negative-spherical-aberration imaging[31] are relatively less commonly used TEM techniques for polarity determination. To the best of our knowledge, very few reports[32] have dealt with the polarity determination of hexagonal AB-stacking QDs using TEM.

In the present study, we have used a modified approach of the NCSI method to determine the polarity of InN QDs grown on a Si (111) substrate by indium droplet epitaxy along with nitridation and further annealing.[33,34] Irrespective of the specimen thickness along the viewing direction, projected atomic columns always appear bright[35] when the image is acquired at an optimum positive defocus keeping the spherical aberration coefficient ($C_s$) of the objective lens at a negative value. This method is termed negative-spherical-aberration imaging (NCSI). For a specific negative $C_s$ value, the optimum positive defocus can be calculated using eqn (8) of the article reported by Jia *et al.*[31] Since the InN QDs of the present study are <10 nm in base diameter and electron beam sensitive in nature,[36-38] it is extremely difficult to set a specific defocus during image acquisition. Therefore, we have used a modified approach of NCSI combined with exit wave reconstruction. At first, the complex-valued exit wave of QDs was reconstructed from a set of focal series images taken at a negative $C_s$ value and finally, an image was generated from the reconstructed exit wave at the calculated optimum positive defocus using the parameters used in the experiment (calculated NCSI image). Besides generating images at the calculated optimum positive defocus, exit wave reconstruction also helps us to minimize image distortion, introduced by the projector lens[39] and delocalization due to coherent aberrations of the objective lens.[40]

## 2. Methods

### 2.1 Synthesis

InN QDs were grown directly on a Si (111) substrate by the technique of Droplet Epitaxy plus a nitridation at low temperature (<100 °C) and later thermal annealing at temperatures below 400 °C. More details on the synthesis can be found elsewhere.[33,34]

### 2.2 TEM sample preparation and imaging

The samples of the materials under study were prepared in cross-sectional disposition and thinned until electron-transparency through a traditional mechanical method comprising grinding, dimpling, polishing and finally Ar$^+$ ion-beam thinning in a GATAN Precision Ion Polishing System (PIPS)-691. By means of double-sided ion-beam etching at small angles (<6°) and low energies (acceleration voltage: <2 kV; beam current <8 µA), the substantial heating of the TEM foils and consequently the introduction of artefacts were avoided.

An FEI-TITAN Cubed Themis 60-300 microscope operating at 200 kV and equipped with an objective lens $C_s$ corrector was used to perform the experiment. During the experiment, the sample was tilted in a way that the InN QDs were viewed along the [11$\bar{2}$0] zone axis, which is perpendicular to the growth direction of the polar InN. The microscope was tuned at an objective lens $C_s$ value of −13.6 µm calculated from the information limit using the amorphous region of the standard calibration sample.

### 2.3 Exit wave reconstruction

A set of 19 images for each QD were acquired at a mean focal step of 2 nm with the defocus values in the range of +20 nm to −16 nm to carry out the exit wave reconstruction using the Full Resolution Wave Reconstruction (FRWR) algorithm[41-43] implemented in Digital Micrograph.[44] The algorithm works by refining an initial guess of the phase iteratively until the reconstructed exit wave becomes self-consistent with the set of focal

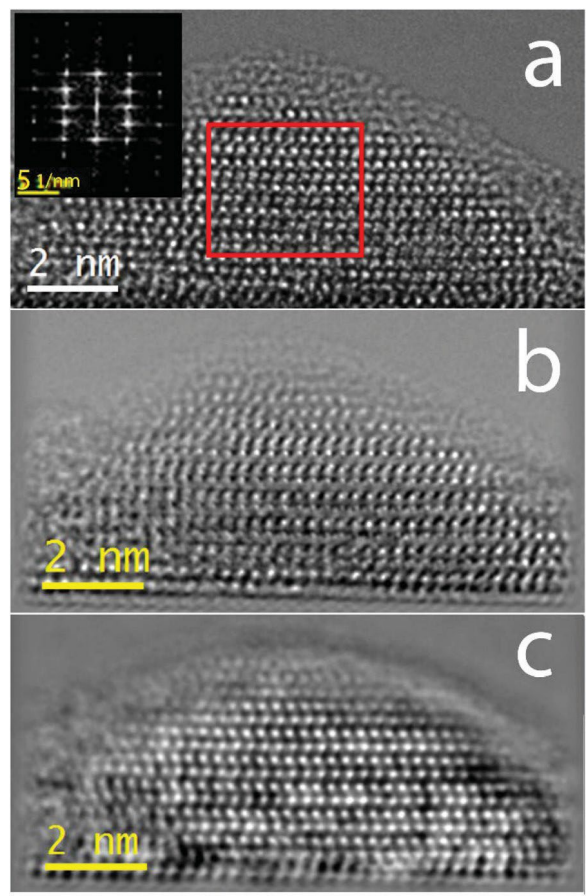

Fig. 1 (a) HRTEM image of the first InN QD from the focal series viewed along the [11$\bar{2}$0] zone axis; the inset shows the power spectrum of the region encompassed by red lines. (b) Retrieved amplitude and (c) phase of the reconstructed exit wave.

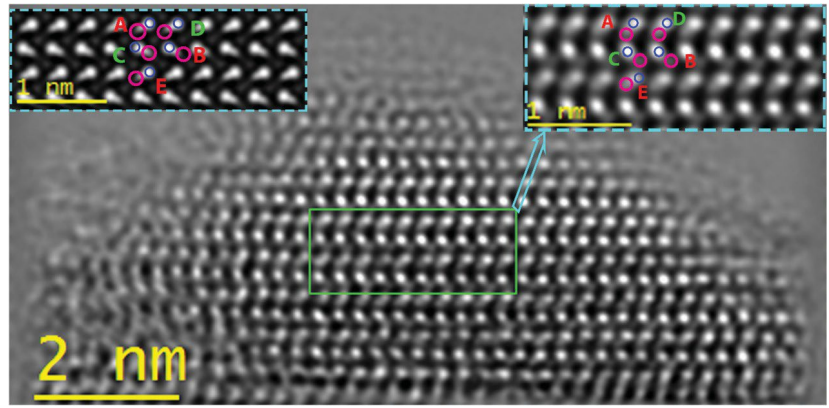

Fig. 2 Calculated NCSI image of the first QD presented in Fig. 1 at 6.7 nm defocus. The right inset shows the zoomed version of the region marked with the green coloured box after background subtraction. The left inset shows the simulated HRTEM image of the N polar InN slab with a thickness of 8 nm along [11$\bar{2}$0] at a defocus of 6.7 nm using a −13.6 μm objective lens $C_s$. Blue circles represent N atoms and pink circles represent In atoms.
The distances between A-B, C-D, and A-E marked in both the insets match with each other.

series images. Two QDs of the same TEM sample were studied in the present work.

### 2.4 DFT calculations and image simulation

Vienna *Ab Initio* Simulation Package (VASP)[45] was used to optimize a wurtzite InN unit cell. The relaxed lattice constants were computed to be $a$ = 3.718 Å and $c$ = 5.381 Å and the internal parameter $u$ = 0.411. The semi-core In 4d electrons were treated as valence electrons. All-electron wave functions and pseudopotentials for the electron-ion interactions were described within the projector-augmented wave (PAW) method.[46] For exchange and correlation (XC) potential, the Perdew, Burke, and Ernzerhof correlation functional (PBE) was used. The Kohn-Sham wave functions were expanded in plane waves up to an energy cutoff of 550 eV with convergence criteria for total energies $1.0 \times 10^{-6}$ eV and forces required to be less than 0.01 eV Å$^{-1}$.

This relaxed wurtzite InN unit cell was used to build a slab structure with both the N and In polarities of thicknesses ranging from 6 nm to 9 nm along the viewing direction [11$\bar{2}$0]. The high-resolution lattice image (HRTEM image) of InN slabs was simulated along [11$\bar{2}$0] using QSTEM[47,48] software. The parameters were kept the same as of the experiment, *e.g.*, 200 kV acceleration voltage, 0.1 convergence angle, −13.6 μm $C_s$, 0.6 eV energy spread ($dE$) of FEG, and 1.4 mm chromatic aberration considering thermal diffuse scattering (TDS).

## 3. Results and discussion

An HRTEM image along the [11$\bar{2}$0] zone axis, selected from a set of focal series images for one QD (first QD) of the present study, is presented in Fig. 1a. The left inset shows the power spectrum of the region encompassed by red lines. Using the maximum spatial frequency ($g_{max}$) measured from this power spectrum, the negative $C_s$ value of the objective lens was calculated[31] which

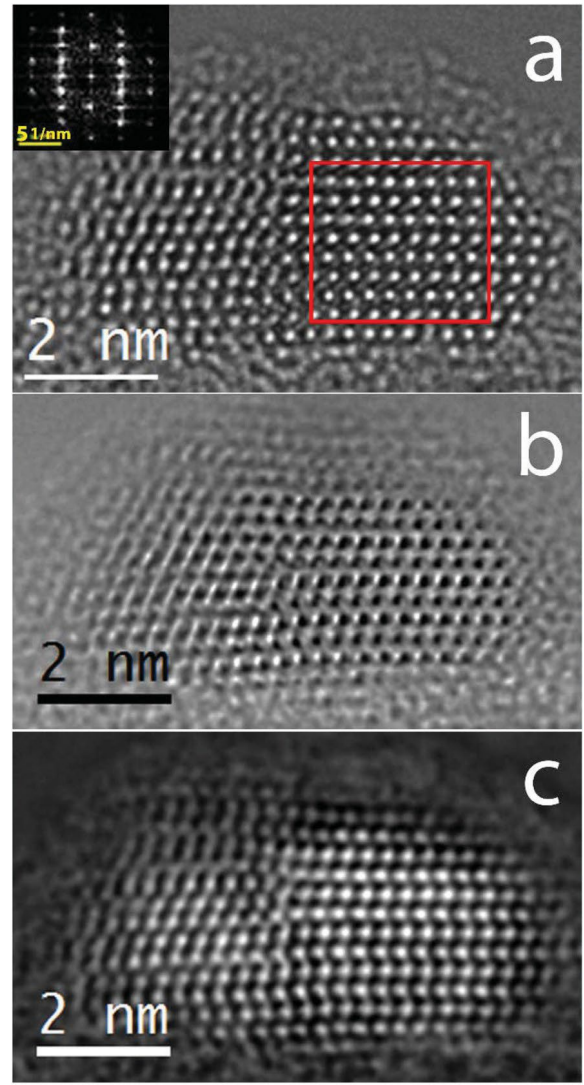

Fig. 3 (a) HRTEM image of the second InN QD from the focal series viewed along the [11$\bar{2}$0] zone axis. (b) Retrieved amplitude and (c) phase of the reconstructed exit wave.

matches well with the one determined using the standard calibration sample. Fig. 1b and c show the amplitude and phase of the reconstructed complex-valued exit wave respectively.

Using the $C_s$ value of −13.6 μm, the optimum positive defocus value was calculated to be 6.7 nm. The calculated NCSI image of the first QD at the defocus of 6.7 nm is presented in Fig. 2. This figure clearly reveals that the calculated NCSI image poses an improved signal to noise ratio leading to better atomic position peak visibility with respect to the single pristine image. Since this QD was surrounded by the leftover of amorphous glue used in cross-sectional TEM specimen preparation, Fig. 2 was further processed with an algorithm[49] to remove the amorphous glue induced artefact. The right inset of Fig. 2 shows a zoomed version of the region marked by the green box after removing amorphous induced artefacts. Since the shape of the QD approached that of a hemisphere,[33] the average horizontal length of the green box encompassed region (∼8 nm) was assumed to be similar to the thickness in the viewing direction. The two InN slab structures of both N and In polarities (using the relaxed wurtzite InN unit cell as presented in Fig. S1 in the ESI †) with a thickness of 8 nm along [11$\bar{2}$0] were used to simulate HRTEM images at the defocus of 6.7 nm with a $C_s$ value of −13.6 μm as described in section 2.4. The simulated HRTEM image of the N polar InN slab is presented in the left inset of Fig. 2 which reveals the same InN dumbbell feature as that of the right inset. At a low index zone axis such as [11$\bar{2}$0], electron channelling through atomic columns containing a higher atomic number element In ($Z$ = 49) is more severe than through atomic columns containing a low atomic number element N ($Z$ = 7). Therefore, in this viewing direction, the feature related to N atomic columns will be larger in diameter than the feature related to In atomic columns. Atomic positions are indicated with blue and pink circles in both the insets where blue represents N atoms and pink represents In atoms. The distances between A-B, C-D, and A-E marked in both the inset images were measured and found to be similar. From this result, we can clearly state that the first QD of the present study is N polar. A model of the N-polar InN structure is provided in Fig. S2 of the ESI. † The simulated HRTEM image of the In-polar InN slab of 8 nm thickness at 6.7 nm defocus is presented in Fig. S3 of the ESI. † In this figure, the 'dumbbell' features are flipped horizontally with respect to the N polar one presented in the left inset of Fig. 2.

Likewise in Fig. 1, an HRTEM image from a focal series of another InN QD (second QD) is presented in Fig. 3, with the corresponding power spectrum in the inset. Its associated amplitude and phase of the reconstructed complex valued exit wave are shown in Fig. 3b and c, respectively.

Fig. 4 presents the calculated NCSI image of the second QD at 6.7 nm defocus. Similar to Fig. 2, the right inset of Fig. 4 presents a zoomed version of the region marked with green after removing amorphous glue induced artefacts. The average horizontal length of this region within the green box is around 6 nm. Likewise in Fig. 2, the simulated HRTEM image of the N polar InN slab of 6 nm thickness at the optimum defocus of 6.7 nm with a $C_s$ of $-13.6$ μm is presented in the left inset which presents a similar InN 'dumbbell' feature to that of the
right inset and all the distances A-B, C-D, and A-E marked in both the insets match with each other. This result confirms that the second QD also possesses N polarity.

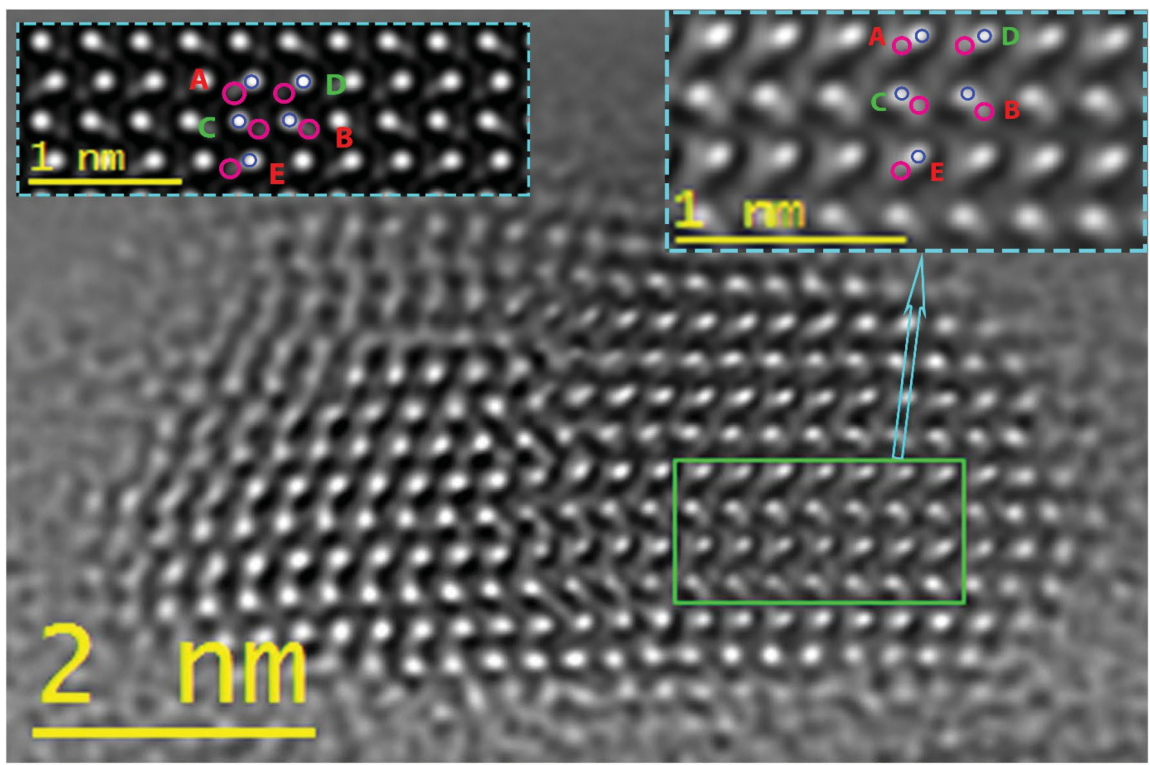

Fig. 4 Calculated NCSI image of the second QD at 6.7 nm defocus. The right inset shows the zoomed version of the region marked with the green coloured box after background subtraction. The left inset shows the simulated HRTEM image of the N polar InN slab with a thickness of 6 nm along [112$\bar{0}$] at a defocus of 6.7 nm using a −13.6 μm objective lens $C_s$. Blue circles represent N atoms and pink circles represent In atoms. The distances between A-B, C-D, and A-E marked in both the insets match with each other.

Fig. 5 presents the simulated HRTEM images of N polar InN slabs with varying thicknesses ranging from 6 to 9 nm with a step of 1 nm along $[11\bar{2}0]$ at defoci of 5 and 9 nm. This figure represents the range of defoci and thicknesses where the N polar InN dumbbell exhibits a feature within HRTEM images which are analogous to the experimental findings.

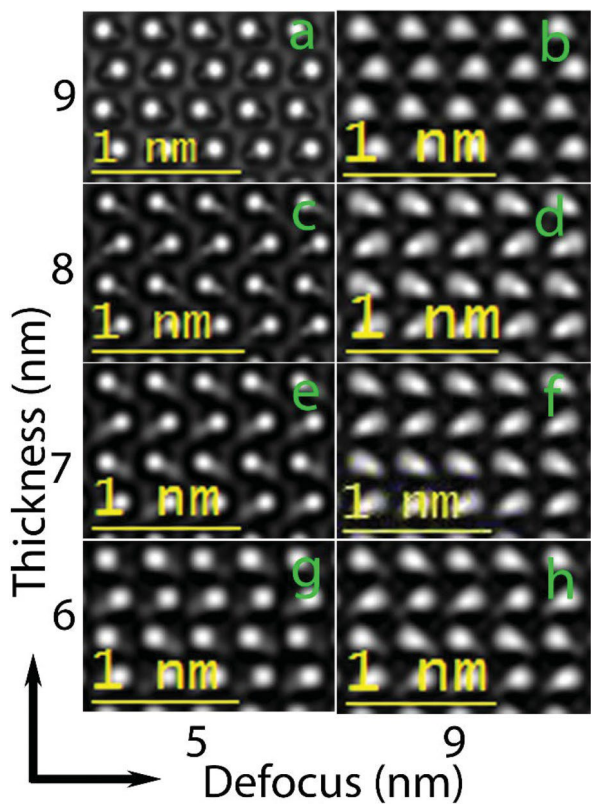

Fig. 5 Simulated thickness – defocus series of the InN structure with N-polarity viewed along the $[11\bar{2}0]$ zone axis with the defocus values of 5 nm and 9 nm and thicknesses of (a-b) 9 nm, (c-d) 8 nm, (e-f) 7 nm and (g-h) 6 nm along the viewing direction.

The STEM imaging of the InN QDs along $[11\bar{2}0]$ was also carried out at an accelerating voltage of 200 kV with a condenser lens $C_s$ value of 0.6 μm and a beam convergence semi-angle of 21 mrad. The image was acquired on an annular detector at a collection semiangle in the range of 12.7 to 24 mrad with a dwell time of 4 μs. The experimental image was compared with the images of 8 nm thick (along $[11\bar{2}0]$) N-polar and In-polar InN slabs simulated using the same experimental parameters. The experimental and simulated STEM images are presented in Fig. S4 of the ESI.† By comparing the experimental data with simulated ones, it is not possible to determine the polarity here due to a lack of spatial resolution. Fig. S5 of the ESI† shows a simulated STEM image with the same parameters as above but with an effective source size of 0.6 Å which reveals the upper limit of effective source size to resolve In and N features separately in the STEM image of a N polar InN slab under present experimental conditions. The effect of source size on image resolution was also discussed earlier for GaN STEM imaging.[50] Therefore, it can be stated that the determination of the polarity of InN QDs using STEM imaging is experimentally challenging.

## 4. Conclusions

A modified approach combining exit wave reconstruction with the NCSI method was used to determine the polarity of InN QDs grown on an Si (111) substrate. In this approach, at first, the complex-valued exit wave was reconstructed using a set of focal series images of an InN QD along $[11\bar{2}0]$ acquired with a negative $C_s$ value of the objective lens, and then, an HRTEM image was generated at the calculated optimum positive defocus from the reconstructed exit wave using the same parameters as of the experiment which poses an improved signal to noise ratio leading to better atomic position peak visibility with respect to the single pristine image. The generated image was compared with HRTEM images simulated at the same optimum defocus using the same experimental parameters from the simulated N and In-polar InN slabs of similar thickness of the experimental imaged region along the viewing direction. Two different QDs within the same TEM specimen were examined and the results revealed that both InN QDs are N polar.

## Author contributions

## Conflicts of interest

The authors declare no competing financial interest.

## Acknowledgements

The financial assistance from the Science and Engineering Research Board (SERB), core research grant project no. EMR/F/2017/001510, is highly acknowledged by the IIT Madras group. This work was partially supported by Spanish MINECO Research Grant Number MAT2015-65120-R and by the University of Cádiz through the "Programa de Fomento e Impulso de la Investigación y la Transferencia" (Proyectos de Investigación-UCA Jóvenes Investigadores PR2016-003 and Puente PR2016-094 and PR2016-042) and the "Programa de ayudas a la realización de Tesis Doctoral del Plan Propio de Investigación y Transferencia" (Contratos predoctorales de Formación de Profesorado Universitario (fpUCA) 2016-060/PU/EPIF-FPU-CT/CP).